\documentclass[prb,superscriptaddress,twocolumn,showpacs,amssymb]{revtex4}

\usepackage{graphicx,bbm,float}

\newcommand{\rutgers}{Department of Physics and Astronomy, Rutgers University,
Piscataway, NJ 08854} 
\newcommand{\anl}{ Center for Nanoscale Materials, Argonne National Laboratory, Argonne IL 60439}
\newcommand{\ucsb} {Materials Department and Materials Research Laboratory, 
University of California, Santa Barbara, CA 93106}

\begin{document}

\preprint{} 

\title{Lattice instabilities in cubic pyrochlore Bi$_2$Ti$_2$O$_7$}

\author{Craig J. Fennie} \affiliation{\anl} 
\author{Ram Seshadri} \affiliation{\ucsb}
\author{Karin M. Rabe} \affiliation{\rutgers}

\date{\today}

\begin{abstract} 
The oxide pyrochlore Bi$_2$Ti$_2$O$_7$ is in some ways analogous to 
perovskite PbTiO$_3$, in that Bi$_2$Ti$_2$O$_7$ has two cations, 
Bi$^{3+}$ and Ti$^{4+}$ in oxidation states that are normally associated with a 
propensity to off-center. However, unlike PbTiO$_3$, 
Bi$_2$Ti$_2$O$_7$ is experimentally observed to remain cubic down to 2\,K,
while the only
observed ionic displacements are local and incoherent. 
Here we report first-principles calculations of the zone-center phonons of the
ordered cubic pyrochlore which reveal several lattice instabilities.
An analysis of the structural energetics suggest that the ordered cubic 
pyrochlore is unstable with respect to distortion towards a 
ferroelectric ground state with a large polarization. Our results
suggest a key role of a frustrated soft polar mode in the dielectric properties 
of bismuth pyrochlores.
\end{abstract}

\pacs{77.84.-s, 61.50.Ks, 63.20.-e}

\maketitle

Chemical disorder has been effectively used as a means of inhibiting 
structural phase transitions and thereby controlling the dielectric properties 
in the ABO$_3$ perovskites. This has engendered the class of relaxor 
ferroelectrics with high and relatively temperature-stable 
dielectric constants.\cite{relaxors,samara} Another path to obtaining new materials with 
inhibited phase transitions is through structural, rather than compositional, 
frustration. This is in analogy with magnetic systems where frustration can be
achieved through substitution (spin glass) or through lattice geometry 
 (spin ice).\cite{ramirez} Recently 
the bismuth pyrochlores have been suggested as potentially belonging to the
class of geometrically frustrated polar materials.\cite{ram.06}

A$_2$B$_2$O$_7$ pyrochlores in the
$\frac12$Bi$_2$O$_3$-ZnO-$\frac12$Nb$_2$O$_5$ (BZN)
system\cite{shrout.ssc.96,shrout.jap.01} have received
considerable attention for capacitor applications due to their
high dielectric constant ($\sim$150), low loss, and absence of any
structural phase transition. At low temperatures, dielectric relaxation 
behavior similar to what is observed in relaxor ferroelectrics is found.
Thin films have been shown to retain the high dielectric constant of 
their bulk counterparts while also displaying a high tunability with 
applied electric field\cite{stemmer.apl.03,stemmer.apl.04} 
suggesting that these materials could easily find applications. 
%
\begin{figure}[b]
\centering \includegraphics[width=8cm]{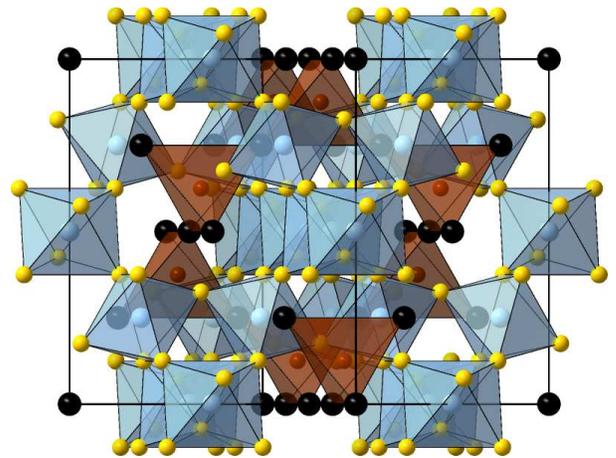}\\
\caption{Color online: A$_2$B$_2$O$_6$O$^\prime$ 
pyrochlore crystal structure showing 
interpenetrating corner-connected networks of O$^\prime$A$_4$ tetrahedra,
and BO$_6$ octahedra. The atom colors are black: A, light blue: B, yellow: O, 
and dark orange: O$^\prime$.}
\label{fig:FIG1} 
\end{figure}
%

The pyrochlore formula can be written A$_2$B$_2$O$_6$O$^\prime$ 
to highlight the view of the cubic $Fd\bar{3}m$ structure as two interpenetrating 
networks; the one of A$_2$O$^\prime$ comprises corner-connected 
O$^\prime$A$_4$ tetrahedra, and the other of B$_2$O$_6$ comprising 
corner connected BO$_6$ octahedra as shown in Fig.~\ref{fig:FIG1}. In 
the ideal pyrochlore structure, the atoms occupy special Wyckoff
positions in the unit cell, A (0,0,0), B (1/2,1/2,1/2), O$^\prime$
(1/8,1/8,1/8),  except O, which is found at (1/8,1/8,$z$).\cite{fd3m_origin} 
In BZN partial substitution of Zn on the Bi A-site occurs and is 
accompanied by the appearance of seemingly uncorrelated and 
random displacements of Bi (and O$^\prime$) atoms off the ideal 
pyrochlore site.\cite{levin_jssc,melot_mrb,withers_jssc}
Even in the absence of cation substitution on the Bi site, pyrochlores 
such as Bi$_2$Ru$_2$O$_7$, Bi$_2$InNbO$_7$, and  Bi$_2$CrTaO$_7$ 
have been shown to accommodate significant static disorder in the 
A$_2$O$^\prime$ network, the cause of which has been suggested 
to originate in the Bi$^{3+}$  lone-pair. The static disorder in this structure 
type is thought to be responsible for both the high and tunable dielectric 
constant as well as the relaxor-like dielectric relaxation yet its precise role 
remains unclear. This has stimulated interest in  
understanding the role of polar phonon modes in the dielectric properties 
of bismuth pyrochlore oxides.\cite{nino.apl.02}

It is clearly advantageous to study a chemically homogeneous system in 
order to separate the role of compositional disorder from geometry. In this 
contribution, we report first-principles calculations of the zone-center 
phonons of the ideal cubic pyrochlore Bi$_2$Ti$_2$O$_7$ as a first step 
towards understanding the origins of the static displacive disorder and its 
possible effects on the polar properties.  Similar to other bismuth
pyrochlores,\cite{levin_jssc} in the bismuth titanate pyrochlore,  
the Bi atoms have been shown to displace incoherently by $\sim$0.4\,\AA\ in 
such a way as to occupy a more general Wyckoff position $(x,x,z)$ while
retaining average cubic symmetry.\cite{hector.jssc.04}
Thin films of the recently prepared compound Bi$_2$Ti$_2$O$_7$ possess
dielectric properties similar to BZN, although they do not seem to be as 
field tunable at room temperature.\cite{stemmer.jap.07}
 Here we show that in the absence of static disorder cubic Bi$_2$Ti$_2$O$_7$ 
has two zone-center lattice instabilities. These couple to produce a low-energy ferroelectric structure with a spontaneous polarization comparable to perovskite
ferroelectrics. We also find that the computed displacement of the Bi ion remarkably similar in magnitude and direction to the measured incoherent distortion. 

First-principles density functional calculations using projector augmented
wave potentials were performed within the local density approximation
as implemented in the \textsc{vasp} program.\cite{VASP,PAW} The 
wavefunctions  were expanded in plane waves up to a kinetic energy cutoff of 
500\,eV. Integrals  over the Brillouiun zone were approximated by sums on a 
$6 \times 6 \times 6$ $\Gamma$-centered $k$-point mesh. Phonon frequencies
and  eigendisplacements were calculated using the direct method
where each symmetry adapted mode\cite{bilbao,isotropy} was moved
by approximately 0.01\,\AA. Born effective charge tensors were
calculated by finite differences of the polarization using the
modern theory of polarization\cite{king-smith.prb.93} as
implemented in \textsc{vasp}.

We performed full structural optimization of Bi$_2$Ti$_2$O$_7$ in
the cubic pyrochlore structure, space group $Fd\bar{3}m$, with
all ions in their ideal Wyckoff positions. In the absence of
incoherent static disorder in the Bi$_2$O$^\prime$ network, we can
investigate the tendency for Bi$_2$Ti$_2$O$_7$ to distort 
coherently by calculating the phonon dispersion. Here we focus on
a subset of such distortions by constructing the dynamical matrix
from finite differences of the Hellmann-Feynman forces at $\vec q=0$
(zone-center) a useful starting point as these modes are
directly relevant to the dielectric properties. 
\begin{table}[t]
\caption{ Infrared-active, Raman-active, and silent phonons frequencies, 
$\omega$ (cm$^{-1}$) of ideal cubic pyrochlore Bi$_2$Ti$_2$O$_7$, 
space group: $Fd\bar{3}m$ (No. 227).}
\begin{ruledtabular}
\begin{tabular}{lcc}
Infrared & Raman & silent  \\ \hline\hline

\begin{tabular}{lr} Mode & $\omega$ \\ 
$T_{1u}$(1) & {\bf $i$98}\\ 
$T_{1u}$(2)&81\\
$T_{1u}$(3)&112\\
$T_{1u}$(4)&229\\
$T_{1u}$(5)&262\\
$T_{1u}$(6)&352\\
$T_{1u}$(7)&464\\ 
\end{tabular}
&
\begin{tabular}{lr} Mode&$\omega$ \\ 
$T_{2g}$(1)& 278\\ 
$E_g$&369\\
$T_{2g}$(2)&414\\
$A_{1g}$&462\\
$T_{2g}$(3)&535\\
$T_{2g}$(4)&462\\
\\ \end{tabular}
&
\begin{tabular}{lr} Mode&$\omega$ \\ 
$E_u$(1)& {\bf $i$100}\\ 
$E_u$(2)&107\\
$E_u$(3)&400\\ \\ \\ \\
\\ \end{tabular}
\end{tabular}
\end{ruledtabular}
\label{table:phonons}
\end{table}
%
%
\noindent Table ~\ref{table:phonons} displays the
calculated infrared, Raman, and selected Silent zone-center phonon 
frequencies and mode assignments.
We find two types of very unstable modes (as indicated by
the large imaginary phonon frequency $\sim i100$\,cm$^{-1}$). The
two-fold  degenerate $E_u$ mode is an antiferrodistortive mode
that breaks inversion symmetry, though without creating a polar
axis. The freezing in of this mode would lead to any of  the
following piezoelectric (but not pyroelectric) space groups\cite{sergienko,isotropy} depending on
anharmonic  contributions: $I\bar{4}m2$, $I4_122$, or $F222$. The
three-fold degenerate $T_{1u}$  mode is a proper ferroelectric
mode which when frozen-in would lead to any one of  the following
space groups: $I\bar{4}2d$, $Ima2$, $R3m$, $Cm$, $Cc$, $P1$. In either
unstable mode, the nature of the eigendisplacements consist, in real-space,
of Bi ions moving perpendicular to the direction of the Bi-O$^\prime$ bond, 
although in both cases a small  cooperative motion of the O network 
is essential as we discuss below. Note this should be contrasted with the nearly unique 
and somewhat ill-described ferroelectric pyrochlore Cd$_2$Nb$_2$O$_7$
where first-principles calculations show that the unstable $T_{1u}$ mode is mostly 
Nb-O character while the $E_u$ mode is stable.\cite{fennie.cd2nb2o7} 
In the $E_u$ case local dipoles are created which cancel when averaged over 
the unit cell while in the $T_{1u}$ case they add to give a spontaneous
polarization. Interestingly, the direction of the Bi ion displacement in both 
unstable modes is in the direction of the experimentally observed static 
displacements. 
\begin{table}[t]
\caption{Possible low-symmetry, ordered structures of the ideal cubic pyrochlore
Bi$_2$Ti$_2$O$_7$, space group $Fd\bar{3}m$. The total energy per formula 
unit $\Delta E$/f.u$.$ (meV) is given relative to the ideal cubic  structure. 
The direction $\mathbf{\hat{P}}$ and magnitude 
$|\mathbf{P}|$ ($\mu$C/cm$^2$) of the spontaneous polarization is 
indicated for the structures that are ferroelectric. }
\begin{ruledtabular}
\begin{tabular}{ccccc}
Space Group	&Irrep 
& $\mathbf{\hat{P}}$&$|\mathbf{P}|$  &$\Delta E$ \\  \hline
$I4_1md$     & $T_{1u}$ &[001] & 40&-92\\
$Ima2$    & $T_{1u}$ &[110] &40 &-107\\
$R3m$    & $T_{1u}$ &[111] &40 &-78\\
$Cm$& $T_{1u}$+$E_u$ &[11$\delta$]  & 20& -240\\
$I\bar{4}m2$  &$E_u$ & - & - &-90\\
$I4_122$  & $E_u$ & -& - &-83\\
\end{tabular}
\end{ruledtabular}
\label{table:energy}
\end{table}

The phonon calculations tell us that at $T=0$, the ideal cubic
pyrochlore Bi$_2$Ti$_2$O$_7$  is indeed unstable and quite
possibly a ferroelectric. To identify which low-symmetry phase 
is the most energetically favorable we performed a series of
structural relaxations within  each of the five highest symmetry
isotropy subgroups consistent with the freezing in of  one
$T_{1u}$ or $E_u$ mode. We also considered one lower 
symmetry structure in space group $Cm$  corresponding to a 
rotation of the polarization along [11$z$]. The results are shown in 
Table~\ref{table:energy}. We found that the lowest energy
structure is ferroelectric, in the polar monoclinic space
group, $Cm$, consistent with the freezing-in of both a [110] T$_{1u}$ mode
and an E$_u$ mode.   For this structure, we calculate a substantial
polarization of $\mathbf{P} \approx$ 20\,$\mu$C\,cm$^{-2}$, and although the
polarization can freely rotate along [11$z$] it remains nearly along
[110].~\cite{polar.direction}

\begin{figure}[t]
\centering \includegraphics[width=5cm]{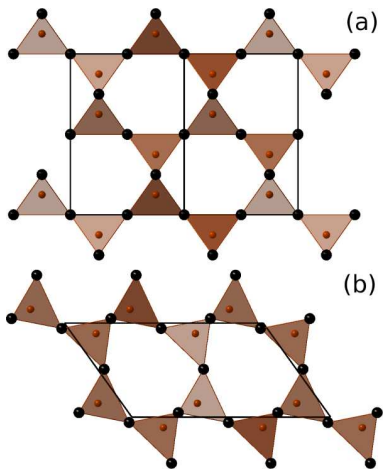}\\
\caption{Color online: (a) Bi$_2$O$^\prime$ network of the cubic Bi$_2$Ti$_2$O$_7$ structure. 
This is the sub-lattice of the structure associated with unstable phonon 
modes. A Projection of the Bi$_2$O$^\prime$ sublattice of the relaxed $Cm$ 
structure. The outline of the monoclinic unit cell is shown.}
\label{fig:FIG3}
\end{figure}

In Fig.~\ref{fig:FIG3}, we compare the Bi$_2$O$^\prime$ networks in the cubic
structure and in the $Cm$ structure. In the $Cm$ structure we find
the average displacement of Bi and O$^\prime$ atoms away from the cubic 
Wyckoff positions
to be 0.37\,\AA\, and 0.13\,\AA\, respectively. Curiously these coherent displacements
are remarkably similar in magnitude to the incoherent Bi and O$^\prime$ displacements
experimentally determined to be 0.43\,\AA\, and 0.16\,\AA\, respectively.
%
%
%
It is tempting to consider this agreement as suggesting what 
is being measured is in some sense an average structure consisting of 
coherently distorted nanodomains formed by the freezing-in of extremely localized, cooperative modes~\cite{morrel.cohen}.
(A signature of this effect can be found from first-principles 
phonon calculations by the appearance of an extremely flat unstable phonon 
branch~\cite{ghosez} at $T=0$; such calculations are underway.)
But we stress, both the T$_{1u}$ and the E$_u$ modes contain significant 
O-character. In the $Cm$ structure for example, O atoms displace from
their high-symmetry Wyckoff position on average by $\approx$ 0.1\AA,
while O-sublattice displacement has not been observed experimentally.

So the question remains as to why incoherent static distortions  in the 
Bi$_2$O$^\prime$ network are observed in neutron studies (from growth
temperatures down to 2\,K) rather than a structural
transition into one of the distorted/ferroelectric phases suggested above. 
Although we do not answer this question here,  it is clear
that these apparent incoherent distortions are not stable in the ground 
state as all of the zero-temperature diagonal elements of the dynamical 
matrix at $\vec q=0$ were positive. This implies for example that the 
Bi and O$^\prime$ sublattices by themselves are stable~\cite{kunes.prb.06,kunes.prb.04}
and highlights the key role played by the O-sublattice in the calculated 
instability. 

The origin of the disorder aside, our first-principles results 
suggests that the key in regard to understanding the dielectric properties 
is a frustrated polar soft mode. In Bi$_2$Ti$_2$O$_7$,  the unstable 
ferroelectric mode and the disorder involve the same ions and a strong 
coupling between the two distinct structural modes is expected.~\cite{yacoby.ssc.97, djsingh.prl.06} 
If we take for a given the presence of 
frozen-in  Bi and O$^\prime$ disorder, one can imagine 
an analogous situation to perovskite relaxor ferroelectrics where intentionally doped chemical disorder frustrates long-range ferroelectric order. 
Further experiments are suggested to elucidate this frustrated
state in Bi$_2$Ti$_2$O$_7$.  In analogy with
the perovskite relaxors ferroelectrics, the effect of cooling in
a strong electric field should be investigated -- can a structural
transition be observed?
If indeed the ground state is ferroelectric and the picture of 
geometric frustration of an infrared-active phonon is valid,
then there should be a significant downshift of spectral weight
observable with optical spectroscopy.\cite{hancock.prl.04} Such a
downshift of spectral weight should also manifest in the
low-temperature heat capacity.\cite{ramirez_nte}

In summary, we have shown that the ideal bismuth titanate
pyrochlore is unstable towards coherent lattice distortions at
$T=0$, the ground state being ferroelectric with Bi and O$^\prime$ ion
displacements off the ideal cubic pyrochlore lattice equal in
magnitude to the measured incoherent distortions. Although it is
still an open question, the results presented here suggest that the
interaction between the static disorder and the soft
ferroelectric mode are responsible for the unusual dielectric
properties of bismuth pyrochlores.

\acknowledgments Useful discussions with M.H.\,Cohen, S.\,Stemmer, 
S.\, Streiffer, and D.H.\,Vanderbilt are gratefully acknowledged. The 
work at Argonne National Lab was supported by US DOE, Office of 
Basic Energy Sciences, under Contract No. DE-AC02-06CH11357.
R.S. has been supported by by the National Science Foundation (DMR 
0449354 and CHE 0434567).
K.M.R. acknowledges support from the Office of Naval Research 
N00014-00-1-0261 and NSF DMR 0507146.
K.M.R. and C.J.F. would also like to acknowledge the
hospitality of the Materials Department at University
of California, Santa Barbara, where part of this work
was carried out.

\clearpage


\begin{thebibliography}{99}
\bibitem{fd3m_origin} Note that the $Fd\bar3m$ space group has two origin
settings.
\bibitem{polar.direction} We performed several 
different relaxations each with the polarization
frozen-in along [11$z$] with different values of z. In all cases, the polarization 
relaxed back to  $\sim$[110].
\end{thebibliography}
\end{document}